\documentclass{ws-procs9x6-cpt25}

\usepackage{slashed}

\begin{document}

\newcommand{\refeq}[1]{(\ref{#1})}
\def\etal {{\it et al.}}

\title{Do Anomalies Break Momentum-Routing Invariance?}

\author{A.R.\ Vieira}

\address{Instituto de Ci\^encias Agr\'arias, Exatas e Biol\'ogicas de Iturama - ICAEBI, Universidade Federal do Tri\^angulo Mineiro - UFTM,\\
Iturama, Minas Gerais 38280-000, Brazil}

\begin{abstract}
The diagrammatic computation of the chiral anomaly 
is associated with momentum-routing invariance breaking. 
This happens because the momentum routing 
in the internal lines of a loop diagram 
is chosen such that the gauge Ward identities hold 
and the chiral Ward identity is broken 
by the finite term 
measured in the pion decay into two photons. 
Since the latter is observable, 
it seems 
that there is a preferred momentum routing 
set by experiments. 
However, 
it is shown in this work 
that the chiral anomaly is momentum-routing invariant. 
This idea is specially important for situations 
in which there are no experiments yet 
to decide on the momentum routing, 
like in supersymmetric theories 
and in frameworks with CPT and Lorentz violation. 
Therefore, 
we resort to momentum-routing invariance 
to find out 
what symmetry is broken 
in a nonminimal CPT- and Lorentz-violating version of quantum electrodynamics.
\end{abstract}

\bodymatter

\section{Introduction}
\label{Sec1}

Although symmetries are one of the main aspects of classical field theories, 
we know that is not possible to maintain all of them at the quantum level. 
The name anomaly is given to a breaking of a classical symmetry 
caused by quantum corrections. 
From the physical point of view, 
we are interested in investigating 
if such anomalies are spurious or indeed physical. 
In the former case, 
it is caused by the regularization scheme, 
that usually breaks symmetries of theories, 
and restoring counter terms are required in the renormalization process. 
In the latter case, 
the anomaly can be measured in an experiment, 
like pion decay into two photons, 
which reveals that chiral symmetry no longer holds 
at the quantum level.\cite{ABJAnomaly}

Since its discovery,\cite{ABJAnomaly} 
the chiral anomaly was derived 
by several different approaches. 
The most common ones, 
usually found in textbooks, 
rely on the fact 
that we should correctly choose the momentum routing 
in the internal lines of the triangle diagrams 
such that the gauge Ward identities hold 
and the chiral Ward identity is broken 
by the term measured in pion decay into two photons. 
Therefore, 
it seems 
that there is a preferred momentum routing 
set by experiment. 
This might lead to the conclusion 
that momentum-routing invariance in Feynman diagrams 
no longer holds 
because it was necessary to choose a specific momentum routing 
in order to get the desired results 
in the Ward identities. 
However, 
the reciprocal is not true, 
i.e., 
anomalies do not imply momentum-routing invariance breaking. 
It is also possible to derive the correct result for the anomaly 
in a momentum-routing invariant way, 
as will be presented in this work.

\section{Momentum-routing invariance of Feynman diagrams}
\label{Sec2}

\begin{figure}
\begin{center}
\includegraphics[width=4in, trim=0mm 70mm 0mm 40mm, clip]{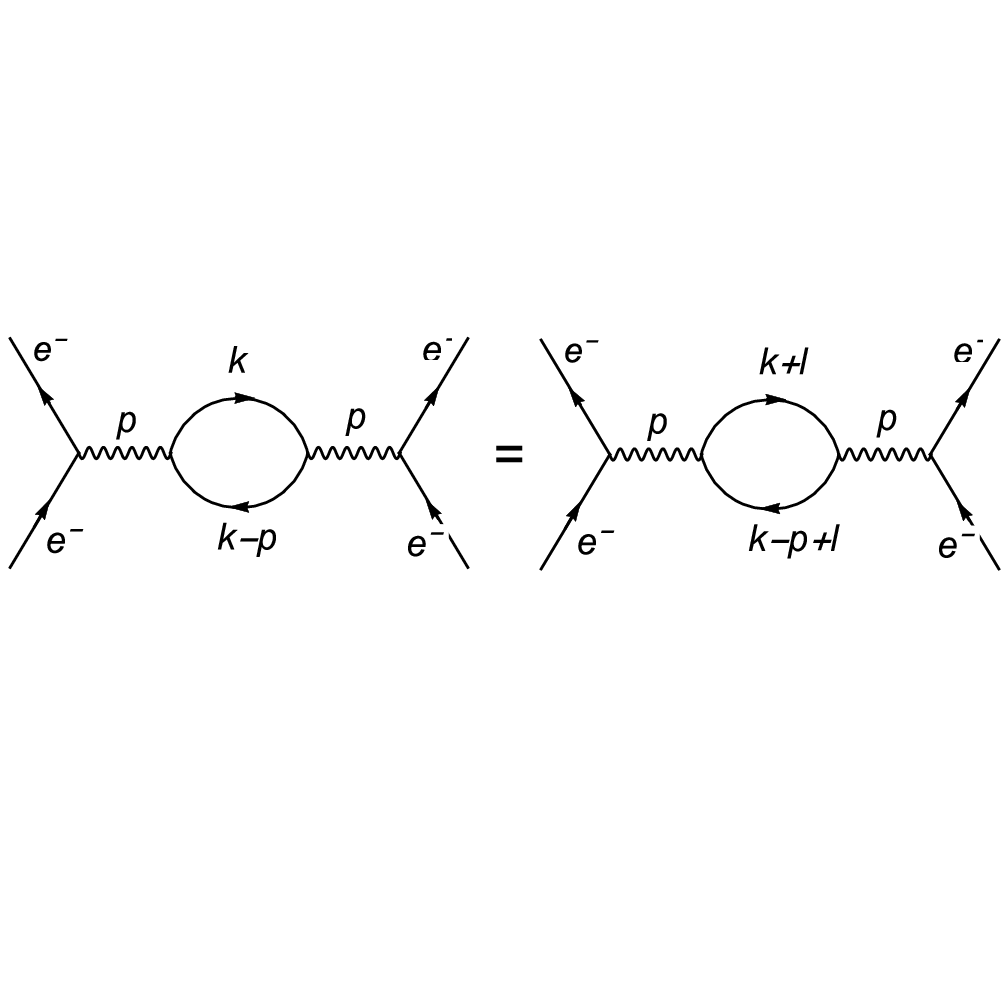}
\end{center}
\caption{Momentum-routing invariance of a QED process.}
\label{fig1}
\end{figure}

Feynman diagrams present the feature of being momentum-routing invariant. 
As a simple example, 
we can consider the one-loop process 
presented in Fig.~\ref{fig1}, 
which is a quantum correction to electron scattering. 
As we see, 
the amplitude of this process 
should not depend on 
whether the momentum of the internal lines is labeled as $k$ or $k+l$ 
because the cross section is an observable. 
However, 
this feature is not usually an issue 
because the most popular regularization scheme is dimensional regularization (DReg),
and this scheme allows for shifts in the integrated momentum. 
In other words,
DReg is momentum-routing invariant 
besides being gauge invariant. 

In abelian and chiral abelian gauge field theories, 
there is an additional reason 
that leads to momentum-routing invariance. 
This invariance is obeyed if and only if 
gauge invariance is as well. 
As a simple example, 
let us consider the gauge Ward identity 
obtained from the vacuum polarization tensor. 
When the amplitude $\Pi^{\mu\nu}(p)$ is contracted with an external momentum, 
we find that this gauge Ward identity 
is nothing but the difference between two tadpoles 
with different momentum routing each. 
This diagrammatic proof is independent of regularization, 
it can be simply achieved algebraically as below 
and it can be generalized to diagrams with more legs and loops:
\begin{align}
p_{\mu}\Pi^{\mu\nu}(p)&=-e^2\int_k {\rm Tr}\left[\slashed{p}\frac{1}{\slashed{k}-\slashed{p}-m}\gamma^{\nu}\frac{1}{\slashed{k}-m}\right]\nonumber\\
&=
-e^2\int_k {\rm Tr}\left[\gamma^{\nu}\frac{1}{\slashed{k}-\slashed{p}-m}\right]
+e^2\int_k {\rm Tr}\left[\gamma^{\nu}\frac{1}{\slashed{k}-m}
\right].
\label{eqMRIVP}
\end{align}
The gauge and momentum-routing invariance relation can be confirmed 
with regularization schemes 
that do not break gauge or momentum routing invariances. 


\section{Momentum-routing invariant chiral-anomaly derivation}
\label{Sec3}

The computation of the chiral anomaly is performed 
for general momentum routings 
with implicit regularization.\cite{AoP} 
In this approach, 
the regularization-dependent content 
resides in the surface terms. 
The corresponding amplitudes of the triangle diagrams are
\begin{align}
T_{\mu\nu\alpha}=&-i\int_k {\rm Tr}\left[\gamma_{\mu}\frac{i}{\slashed{k}+\slashed{k}_1-m}\gamma_{\nu}\frac{i}{\slashed{k}+\slashed{k}_2-m}
\gamma_{\alpha}\gamma^5 \frac{i}{\slashed{k}+\slashed{k}_3-m}\right] \nonumber\\
&+(\mu \leftrightarrow \nu, p\leftrightarrow q).
\label{AVV}
\end{align}

General momentum routings are constrained 
by energy-momentum conservation at each vertex: 
$k_2-k_3= p+q$, 
$k_1-k_3= p$, 
and $k_2-k_1= q$. 
This allows us to parameterize general momentum routings as 
$k_1= \alpha p+(\beta-1) q$, 
$k_2= \alpha p+\beta q$, 
and $k_3= (\alpha-1) p+(\beta-1) q$, 
where $\alpha$ and $\beta$ are real numbers. 
After computing the diagrams and applying the respective external momentum 
in order to obtain the Ward identities, 
we find the following results:
\begin{align}
	p_{\mu}T^{\mu \nu \alpha}&=-4i\upsilon_0 (\alpha-\beta-1)\epsilon^{\alpha \nu \beta \lambda}p_{\beta}q_{\lambda},\nonumber\\
	q_{\nu}T^{\mu \nu \alpha}&=4i\upsilon_0 (\alpha-\beta-1)\epsilon^{\alpha \mu \beta \lambda}p_{\beta}q_{\lambda},\nonumber\\
	(p+q)_{\alpha}T^{\mu \nu \alpha}&=2m T_5^{\mu \nu}+8i\upsilon_0 (\alpha-\beta-1)\epsilon^{\mu \nu \beta \lambda}p_{\beta}q_{\lambda}-\tfrac
	{1}{2\pi^2}\epsilon^{\mu \nu \beta \lambda}p_{\beta}q_{\lambda},
	\label{WI}
\end{align}
where $T_5^{\mu \nu}$ is the usual vector-vector-pseudoscalar triangle.

It is easy to see in Eq.~(\ref{WI}) 
that if the surface term $\upsilon_0$ is zero, 
we get the correct result of the chiral anomaly 
$-\frac{1}{2\pi^2}\epsilon^{\mu \nu \beta \lambda}p_{\beta}q_{\lambda}$ 
and gauge symmetry holds in the first two equations. 
Besides, 
this result is momentum-routing invariant 
since $\alpha$ and $\beta$ can be any real numbers for a null surface term.

\section{The chiral anomaly in a $d=5$ Lorentz-violating QED}

We can resort to momentum-routing invariance 
to find out 
what symmetry is broken in situations 
where there are no experiments yet 
to fix the momentum-routing ambiguity.\cite{AoP} 
This would be the case in supersymmetric theories 
or frameworks with CPT and Lorentz violation. 
We consider the mass dimension $d=5$ operator 
in the fermion sector $-\frac{1}{2}(a^{(5)}_F)^{\mu\alpha\beta}\overline{\psi}\gamma_{\mu}F_{\alpha\beta}\psi$ 
in the Lagrangian $\mathcal{L}^{(5)}_{\psi F}$ 
of Table I in Ref.~[\refcite{Zonghao}], 
where $(a^{(5)}_F)^{\mu\alpha\beta}$ 
is a set of CPT- and Lorentz-violating coefficients. 
It can be rewritten as 
$-(a^{(5)}_F)^{\mu\alpha\beta}\overline{\psi}\gamma_{\mu}\partial_{\alpha}A_{\beta}\psi$ 
due to antisymmetry in $\alpha , \beta$.

\begin{figure}[!h]
\centering
\centering
\includegraphics[trim=0mm 160mm 35mm 0mm, scale=0.35]{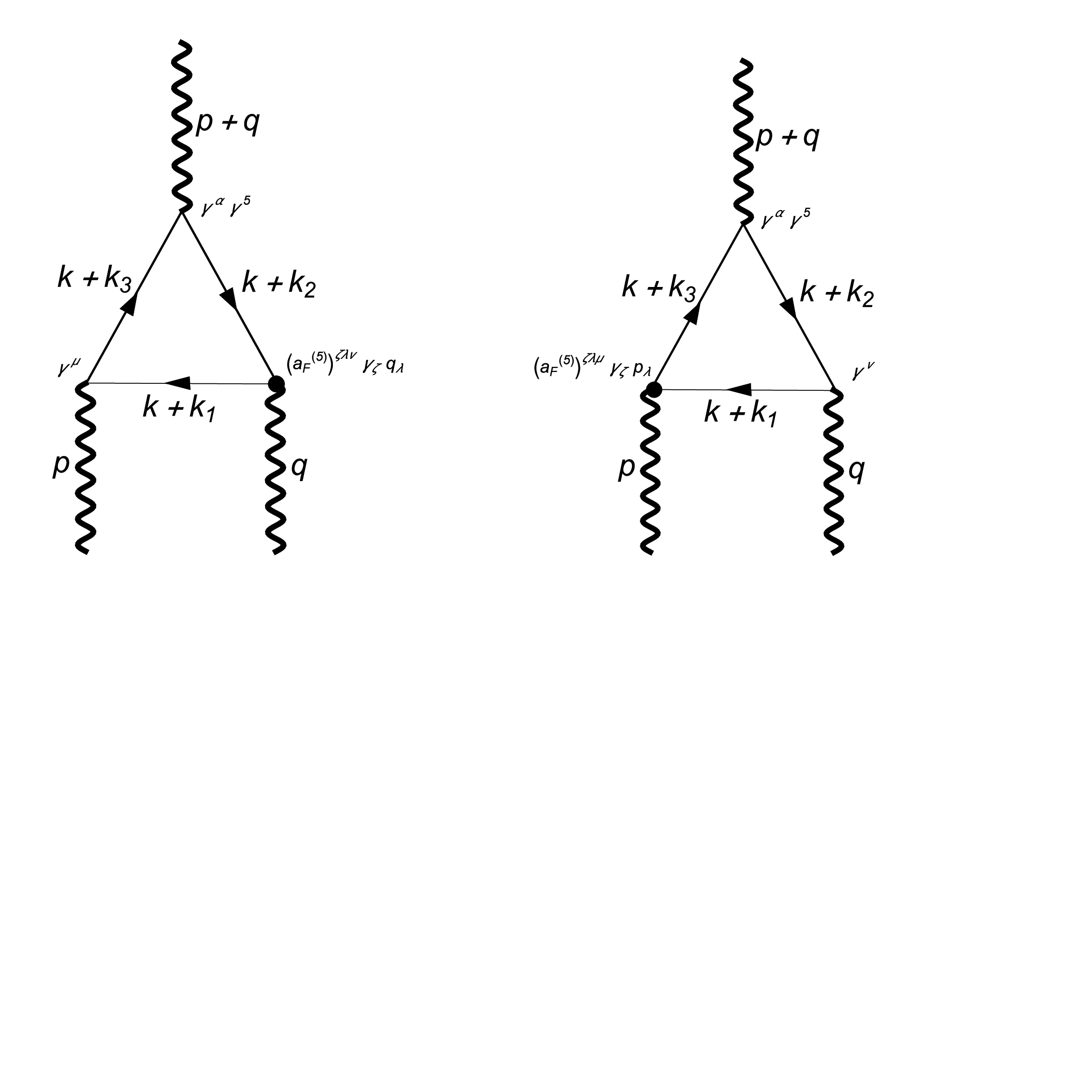}
\caption{Triangle diagrams in a QED with nonminimal Lorentz violation.}
\label{fig4}
\end{figure}

The triangle diagrams are depicted in Fig.~\ref{fig4}. 
As in the usual case, 
this contribution must be summed 
with the crossed diagrams. 
Its corresponding amplitude can be written as 
\begin{align}
T^{\rm LV}_{\mu\nu\alpha}=&-i\int_k {\rm Tr}\left[\gamma_{\mu}\frac{i}{\slashed{k}+\slashed{k}_1-m}(a_F^{(5)})_{\zeta\lambda\nu}\gamma^{\zeta}q^{\lambda}\frac{i}{\slashed{k}+\slashed{k}_2-m}\gamma_{\alpha}\gamma^5 \frac{i}{\slashed{k}+\slashed{k}_3-m}\right] \nonumber\\
&-i\int_k {\rm Tr}\left[(a_F^{(5)})^{\zeta\lambda}_{\ \ \ \mu}\gamma_{\zeta}p_{\lambda}\frac{i}{\slashed{k}+\slashed{k}_1-m}\gamma_{\nu}\frac{i}{\slashed{k}+\slashed{k}_2-m}\gamma_{\alpha}\gamma^5 \frac{i}{\slashed{k}+\slashed{k}_3-m}\right] \nonumber\\
&+(\mu \leftrightarrow \nu, p\leftrightarrow q).
\label{AVVLV}
\end{align}

The parameterization for the arbitrary routing of $k_i$ 
is the same as in the previous section. 
Now, 
if we compute the Ward identities 
we find, 
as expected, 
that gauge symmetry is broken by the arbitrary surface term $\upsilon_0$, 
as presented below for the massless limit
\begin{align}
&p_{\mu}T^{LV\mu\nu\alpha}=-8i\beta\upsilon_0 (a^{(5)}_{F})^{\ \ \ q\nu}_{\zeta}\epsilon^{q\alpha\zeta p},\nonumber\\
&q_{\nu}T^{LV\mu\nu\alpha}=8i(1-\alpha )\upsilon_0(a^{(5)}_{F})^{\ \ \ p\mu}_{\zeta}\epsilon^{q\alpha\zeta p},\nonumber\\
(p_{\alpha}+q_{\alpha})T^{LV\mu\nu\alpha}&=8i(\alpha-\beta-1 )\upsilon_0((a^{(5)}_{F})^{\ \ \ q\nu}_{\zeta}\epsilon^{q\zeta\mu p}-(a^{(5)}_{F})^{\ p\mu}_{\zeta}\epsilon^{q\zeta\nu p})\nonumber\\
&-\frac{1}{2\pi^2}((a^{(5)}_{F})^{\ \ \ q\nu}_{\zeta}\epsilon^{q\zeta\mu p}+(a^{(5)}_{F})^{\ \ \ p\mu}_{\zeta}\epsilon^{p\zeta\nu q}),
\label{IWsLV}
\end{align}
where $(a^{(5)}_F)^{\alpha p\mu}\equiv (a^{(5)}_F)^{\alpha \lambda\mu}p_{\lambda}$ and $\epsilon^{\alpha\lambda\nu p}\equiv\epsilon^{\alpha\lambda\nu 
\sigma}p_{\sigma}$.

In this case, 
there is no particle process, 
such as pion decay, 
where the breaking of chiral symmetry 
provided by Lorentz violation at tree level 
is measured. 
So, 
the ambiguity in the momentum routing 
can not be solved 
by means of an experiment. 
However, 
since we showed in the previous section 
that momentum-routing invariance is not broken 
in the usual chiral anomaly, 
we can resort to this invariance in Eqs.~(\ref{IWsLV}). 
We see 
that the results of the Ward identities 
are momentum-routing invariant 
if the surface term $\upsilon_0$ is null. 
As a consequence, 
we have gauge symmetry and a chiral anomaly 
induced by CPT and Lorentz violation at the quantum level. 
It is noteworthy 
that some choices of momentum routing 
such as $\alpha=0$ and $\beta=-1$ 
are compatible with the Ward identities of the Lorentz-invariant situation measured by experiments, 
i. e., 
gauge symmetry holding with no chiral symmetry, 
as we can see from Eqs.~(\ref{WI}). 
Nevertheless, 
this choice of momentum routing 
breaks both gauge and chiral symmetry 
in the Lorentz-violating situation 
as we can see in the Ward identities~(\ref{IWsLV}). 
We can instead alternatively choose a null value of the surface term, 
so that the Ward identities of both cases 
are compatible with momentum-routing invariance. 
In this case, 
we would have in both situations 
gauge symmetry and the breaking of chiral symmetry.

\end{document}